\documentclass{PoS}

\newcommand{\beq}{\begin{equation}}
\newcommand{\eeq}{\end{equation}}
\newcommand{\bea}{\begin{eqnarray}}
\newcommand{\eea}{\end{eqnarray}}
\newcommand{\ba}{\begin{align}}
\newcommand{\ea}{\end{align}}
\newcommand{\bfig}{\begin{figure}}
\newcommand{\efig}{\end{figure}}

\newcommand{\D}{\displaystyle}

\newcommand{\tin}{t_{\rm in}}
\newcommand{\la}{\langle}
\newcommand{\ra}{\rangle}

\title{Bounds on the spacelike pion electromagnetic form factor from
  analyticity and unitarity}

\ShortTitle{Bounds on the spacelike pion electromagnetic form factor}

\author{\speaker{I. Sentitemsu Imsong}\\
        Centre for High Energy Physics, Indian Institute of Science,\\
        Bangalore 560012, India\\
        E-mail: \email{senti@cts.iisc.ernet.in}}

\author{B.Ananthanarayan\\
         Centre for High Energy Physics,
	Indian Institute of Science, Bangalore 560 012, India\\
        E-mail: \email{anant@cts.iisc.ernet.in}}
 
 \author{Irinel Caprini\\
         National Institute of Physics and Nuclear Engineering\\
 	POB MG 6, Bucharest, R-76900, Romania\\
        E-mail: \email{caprini@theory.nipne.ro}}

\abstract{We use the recently measured accurate BaBaR data on the
 modulus of the pion electromagnetic form factor,  $F_\pi(t)$, up to an
 energy of 3 GeV, the $I=1$ $P$-wave phase of the $\pi\pi$ scattering
 amplitude up to the $\omega-\pi$ threshold, the pion charge radius known
 from Chiral Perturbation Theory, and the recently measured JLAB value of
 $F_\pi$ in the spacelike region at $t=-2.45 {\rm GeV}^2$ as inputs in a
 formalism that leads to   bounds on $F_\pi$  in the intermediate spacelike
 region.  We compare our constraints with experimental data and with perturbative 
 QCD along with the results of several theoretical models for the non-perturbative contributions
 proposed in the literature.}

\FullConference{Sixth International Conference on Quarks and Nuclear Physics,\\
		April 16-20, 2012\\
		Ecole Polytechnique, Palaiseau, Paris}

\begin{document}

\section{Introduction}
The spacelike pion electromagnetic form factor $F_\pi(t)$ is of interest
for studying the onset of perturbative QCD (pQCD). At leading order (LO),
the expression for $F_\pi(t)$, calculated in pQCD is given by \cite{Farrar:1979aw,Lepage:1979zb,Efremov:1979qk,pQCD3}
\beq\label{eq:qcd1}
  F^{\rm LO}_{\rm pert}(-Q^2)= \frac{8 \pi f_\pi^2\alpha_s(\mu^2 )}{Q^2},  \quad \quad \quad t(=-Q^2)<0
 \eeq
while  the next-to-leading order (NLO) correction with the ${\overline{\rm MS}}$-renormalization 
scheme and asymptotic DAs reads \cite{Melic:1999mx}
\beq\label{eq:qcd2}
F^{\rm NLO}_{\rm pert}(-Q^2)= \frac{8 f_\pi^2\alpha_s^2(\mu^2)}{Q^2}\,\left[\frac{\beta_0}{4}\left( \ln \frac{\mu^2}{Q^2}+\frac{14}{3}\right) - 3.92\right],
 \eeq 
where $f_\pi= 130.4$ MeV is the pion decay constant and $\alpha_s(\mu^2)$ the  strong  coupling at 
the renormalization scale $\mu^2$. The quantity $\beta_0=11-2 n_f/3$ denotes the first coefficient in the perturbative 
expansion of the $\beta$-function, $n_f$ being the number of active flavors. 

There is an interplay of perturbative with soft, nonperturbative processes especially in the intermediate
$Q^2$ region as a result of which the asymptotic regime sets in quite slowly in the case of the pion form factor.
Therefore, it remains an open question as to at what value of $Q^2$ do the nonperturbative contributions
become negligible so that the perturbative QCD description of the form factor becomes reliable.
Several nonperturbative approaches have been proposed for the study of the spacelike pion form factor 
\cite{Braun:1999uj, Radyushkin:2001, Bakulev:2004cu, Brodsky:2007hb,Grigoryan:2007, Braguta:2007fj, Bakulev:2009ib}.
On the experimental side, measurements of the spacelike pion form factor at various energies are available,
the most recent data coming from the $JLAB$, \cite{Horn, Huber}. Further, we now have  
more accurate information on the phase \cite{GarciaMartin:2011cn}
and modulus \cite{BABAR} of the pion form factor on the unitarity cut. In this paper, we
perform an analytic continuation from the timelike to the spacelike 
region using in a most conservative way the available information on the phase and modulus on 
the unitarity cut, and also the spacelike information available. Using a mathematical formalism discussed in
\cite{IC, Abbas:2010EPJA}, we find stringent upper and lower bounds at different values of spacelike momenta, 
which allows us to find a lower limit for the onset of the QCD perturbative behavior \cite{Ananthanarayan:2012tn}.

 \section{Formalism}

Our formalism requires the knowledge of the phase below $t_{\rm in}$  
and an integral over the modulus squared from $t_{\rm in}$ to $\infty$.
We relate the phase of pion the form factor with that of the
associated $\pi\pi$ scattering amplitude via the Fermi Watson theorem. 
In this case, we consider the relation
\beq\label{eq:watson}
{\rm Arg} [F(t+i\epsilon)]=\delta_1^1(t), \quad\quad  4 M_\pi^2 \leq t \leq \tin,
\eeq
where $\delta_1^1(t)$ is the phase shift of the $P$ wave of $\pi\pi$ elastic scattering and
$\tin=(M_\pi+M_\omega)^2$ is the upper limit of the elastic region,  which corresponds to 
the first important inelastic threshold due to the $\omega\pi$ pair.
We use the recent experimental data on the modulus up to $\sqrt{t}=3\, {\rm GeV}$ \cite{BABAR}. 
Above this energy, we make conservative assumptions and obtain a rather accurate estimate of an integral of 
modulus  squared from $\tin$ to infinity. More precisely, we assume the following condition,
 \beq\label{eq:L2}
 \D\frac{1}{\pi} \int_{\tin}^{\infty} dt \rho(t) |F_\pi(t)|^2 = I,
 \eeq
where $\rho(t)$ is a suitable positive-definite weight, for which the integral converges,  
and the number $I$  can be estimated with sufficient precision. The optimal procedure is to vary $\rho(t)$ over 
a suitable admissible class and take the best result. In principle, a large 
class of positive weights, leading to a convergent integral for $|F_\pi(t)|$ compatible with the asymptotic 
behavior  (\ref{eq:qcd1}) of the pion form factor, can be adopted. In our 
calculations, we consider an expression of the form 
\beq \label{eq:rhogeneric}
\rho_a(t) = \frac{1}{t^a}, \quad \quad  0\leq a \leq 2
\eeq

We use additional information inside the analyticity domain namely the normalization $F_\pi(0)=1$
and the pion charge radius, $F_\pi'(0)= \la r_\pi^2 \ra/6$,
with $\la r_\pi^2 \ra$ varied within reasonable limits \cite{Masjuan,Colangelo:2004}, 
and the values of the form factor at some spacelike values $F(t_n)$ where $t_n<0$ \cite{Horn, Huber}.
In this paper, we derive rigorous upper and lower bounds on $F_\pi(t)$ in the region $t<0$, for 
functions $F_\pi(t)$ belonging to the class of real analytic functions in the $t$-plane cut for $t>4 M_\pi^2$
defined by all the inputs specified. 

For solving the problem, we apply a standard mathematical method discussed in detail in \cite{IC,Abbas:2010EPJA}. 
We transform our problem via a conformal map, cast the
integral equation into a canonical form and derive a determinant (see ref. \cite{Ananthanarayan:2012tn}
for more details) which is central to our investigations for obtaining bounds on $F_\pi(t)$ in
the spacelike region.

 \section{Inputs} \label{sec:inputs}

In the elastic region $t\leq \tin$ we use the phase shift parameterization 
determined recently with high precision from Roy equations applied 
to the $\pi\pi$ elastic amplitude given in \cite{GarciaMartin:2011cn} (see ref.\cite{Ananthanarayan:2012tn}
for more details). Above $\tin$ we choose $\delta$ as a continuous function,
sufficiently smooth which approaches asymptotically $\pi$. 
The results are independent of the choice of $\delta(t)$ above $\tin$, as
discussed in detail in \cite{Abbas:2010EPJA}.
For the calculation of the integral defined in (\ref{eq:L2}), we
use the BaBar data \cite{BABAR} from $\sqrt{\tin}=0.917$ GeV up to $\sqrt{t}=3$ GeV,  and 
have taken a constant value for the modulus in the range $3\,\, {\rm GeV} \leq \sqrt{t} \leq 20\,\, {\rm GeV}$,  
continued with a $1/t$ decrease above 20 GeV. It may be noted that our estimates are based on
very conservative assumptions of the input quantities which makes our procedure very robust.
 In our analysis, we consider the weights of the form given in (\ref{eq:rhogeneric}). 
The values of $I$ corresponding to several choices of the parameter $a$ are given in Table 1 of 
ref. \cite{Ananthanarayan:2012tn}, where the uncertainties are due to the BaBar experimental errors.  
We find from our analysis that the best results come from an optimal
choice of the weight corresponding to $\rho_{1/2}(t)$.
We use also as input,
\beq\label{eq:r2}
  \la r_\pi^2 \ra =0.43 \pm 0.01 \,\mbox{fm}^2, \quad \quad \quad F( -2.45 {\rm GeV}^2)= 0.167 \pm 0.010_{-0.007}^{+0.013},
\eeq
respectively for the pion charge radius \cite{Masjuan,Colangelo:2004} and the spacelike datum \cite{Horn,Huber}.

 \section{Results}\label{sec:results}

\begin{figure}[htb]
 \vspace{0.35cm}
\begin{center}
  \includegraphics[width = 8.cm]{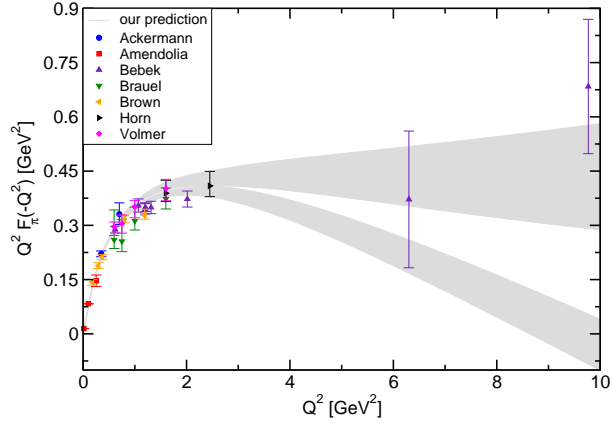}
 \caption{Allowed domain obtained with the weight $\rho_{1/2}(t)$  compared with several sets of experimental data.}
 \label{fig:fig6}
 	\end{center}
 \end{figure}

\begin{figure}[htb]
 \vspace{0.35cm}
\begin{center}
  \includegraphics[width = 8.cm]{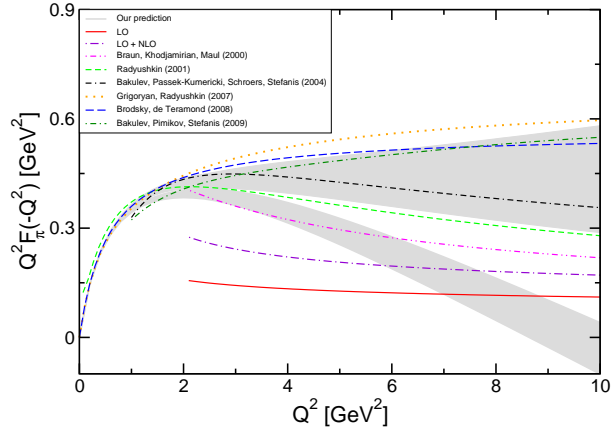}
 \caption{Comparison of the bounds for the weight $\rho_{1/2}(t)$, with perturbative QCD and several nonperturbative models.}
 \label{fig:fig8}
 	\end{center}
 \end{figure}

The main results emerging from our analysis are presented in Figs. \ref{fig:fig6}
and \ref{fig:fig8}. In the figures, the white band corresponds
to the bound obtained by using only the central values of the inputs while
the grey bands are obtained from the errors associated with the inputs.
The error bands have been obtained by adding quadratically the errors produced 
by the variation of the phase, the charge radius, the integral $I$ for $a=1/2$ 
from Table 1 of ref. \cite{Ananthanarayan:2012tn}, 
and the spacelike datum. We find that the greatest contribution to the size 
of the grey domain is the experimental uncertainty associated with the spacelike value  (\ref{eq:r2}).
In Fig. \ref{fig:fig6}, we compare our constraints with some of the 
data available from experiments (see \cite{Ananthanarayan:2012tn} for references). 
We find that at lower values of $Q^2$ most of the data are 
consistent with the narrow allowed band predicted by our analysis. 
In Fig. \ref{fig:fig8}, we compare our allowed domain with the pQCD predictions both at LO and NLO 
and with various nonperturbative models. The perturbative prediction to NLO is sensitive to 
the choice of the renormalization scale, and also of the factorization scale in the case when 
pion DAs different from the asymptotic ones are used in the calculation \cite{Melic:1999mx,Brodsky:1998} . 
Several prescriptions for scale setting have been adopted, but there is no general consensus on the issue.
For illustration, in  Fig. \ref{fig:fig8}  we show the sum of the LO and NLO terms  (\ref{eq:qcd1}) and 
(\ref{eq:qcd2}),  obtained with the  scale $\mu^2=Q^2$ and the one loop coupling.   
This curve is compatible with our bounds enlarged by errors only for $Q^2>6\,\, {\rm GeV}^2$.
We show also several  nonperturbative models proposed in the literature 
for the spacelike form factor at intermediate region \cite{Braun:1999uj, Radyushkin:2001,
Bakulev:2004cu, Brodsky:2007hb,Grigoryan:2007, Braguta:2007fj, Bakulev:2009ib}.
In Ref. \cite{Braun:1999uj}, the authors applied light-cone QCD sum rules and parametrized with a 
simple expression the nonperturbative  correction, to be added to the LO+NLO perturbative prediction 
in the region $1<Q^2<15\,{\rm GeV}^2$. We show the sum of the soft correction and the 
perturbative QCD prediction to NLO,  evaluated at a scale $\mu^2=0.5\, Q^2 + M^2$ with $M^2=1\,{\rm GeV}^2$ 
as argued in \cite{Braun:1999uj}. The model is quite compatible with our bounds, the corresponding curve
being inside the small white inner domain for $Q^2>6\,\, {\rm GeV}^2$.
The model based on local duality \cite{Radyushkin:2001} is also consistent with the allowed 
domain derived here for $Q^2> 1\,{\rm GeV}^2$. We mention that this model, proposed in \cite{Nesterenko:1982}, was 
recently developed by several authors  \cite{Braguta:2007fj}.  The other models shown in  Fig. 
\ref{fig:fig8} are consistent with the bounds derived by us at low $Q^2$, but are at the upper limit 
of the allowed domain at higher $Q^2$. The agreement is somewhat better for the model discussed in 
\cite{Bakulev:2004cu}, which is a LO+NLO perturbative calculation using nonasymptotic pion DAs evolved 
to NLO, with a modification of the QCD coupling by the so-called analytic perturbation theory. 
The AdS/QCD model considered in \cite{Brodsky:2007hb} is in fact a simple dipole interpolation, 
which is valid at low energies but seems to overestimate the form factor
at larger momenta. The same remark holds for the models 
discussed in \cite{Grigoryan:2007} and \cite{Bakulev:2009ib},  based on QCD sum rules with nonlocal 
condensates, and the chiral limit of the hard-wall AdS/QCD approach, respectively.

\bigskip

\section{Conclusion}
We have derived upper and lower bounds on the pion electromagnetic form factor along the spacelike 
axis, by exploiting in a conservative way the precise information on the phase and modulus 
of the timelike axis as well as the available spacelike data. We have used the method of 
analytic continuation to obtain information on the spacelike region of the pion form factor.
Using the weight $\rho_{1/2}(t)$ which is the optimal choice as discussed in Sec. \ref{sec:inputs}, we have
obtained upper and lower bounds enlarged by errors associated with the various inputs entering
our analysis. From Fig. \ref{fig:fig8}, we can conclude that 
perturbative QCD to LO is excluded  for $Q^2< 7\,\,{\rm GeV}^2$,  and perturbative QCD 
to NLO is excluded  for $Q^2< 6\,\,{\rm GeV}^2$, respectively. If we restrict to the inner white allowed 
domain obtained with the central values of the input, the exclusion regions  become 
$Q^2< 9\,\,{\rm GeV}^2$ and  $Q^2< 8\,\,{\rm GeV}^2$, respectively. Among the theoretical models, the 
light-cone QCD sum rules \cite{Braun:1999uj} and the local quark-hadron duality model \cite{Radyushkin:2001}  
are consistent with the allowed domain derived here for a large energy interval, while the remaining 
models are consistent with the bounds at low energies, but seem to predict too high values at higher $Q^2$.
To increase the strength of the predictions, a reduction of the 
grey bands produced by the uncertainties of the input is desirable. 
As mentioned in Sec.\ref{sec:results}, the biggest contribution to the error band comes from 
the experimental errors associated with the spacelike datum (\ref{eq:r2}).  
As such, in order to increase the predictive power of our formalism, 
more accurate data at a few spacelike points, particularly at larger values of $Q^2$, 
would be very useful.

\end{document}